\documentstyle[prl,aps,epsf,twocolumn]{revtex}

\def\cZ{{\cal Z}}
\def\cP{{\cal P}}
\def\cG{{\cal G}}
\def\cL{{\cal L}}
\def\cF{{\cal F}}
\def\cT{{\cal T}}
\def\G{\Gamma}
\def\ba{\begin{array}}
\def\ea{\end{array}}
\def\be{\begin{equation}}
\def\ee{\end{equation}}
\def\bea{\begin{eqnarray}}
\def\eea{\end{eqnarray}}
\def\6{\partial}
\def\5{\bar }
\def\7{\tilde }
\def\8{\hat }

\def\nn{\nonumber}
\def\hmu{\hat{\mu}}
\def\hnu{\hat{\nu}}

\newcommand{\slPi}{/ {\hskip-0.27cm{\Pi}}}

\font\mybb=msbm10 at 10pt
\def\bb#1{\hbox{\mybb#1}}
\def\bZ {\bb{Z}}
\def\bR {\bb{R}}

\begin{document}

\title{
{\normalsize\begin{flushright}
UB--ECM--PF--00/11\\
WIS/21/00-OCT-DPP\\
hep-th/0010242\\[1.5ex]
\end{flushright}}
Automorphisms as brane non-local transformations
}

\author{
Joan Sim\'on }

\address{
Departament ECM,
Facultat de F\'{\i}sica,
Universitat de Barcelona and 
Institut de F\'{\i}sica d'Altes Energies \\
Av. Diagonal 647,
E-08028 Barcelona, Spain \\
and \\
Department of Particle Physics, The Weizmann Institute of Science \\
2 Herzl Streeet, 76100 Rehovot, Israel.\\
E-mail: jsimon@weizmann.ac.il.\\[1.5ex]
\begin{minipage}{14cm}\rm\quad
The relation among spacetime supersymmetry algebras and superbrane actions
is further explored. It is proved that $SL(2,\bR)$ belongs to the automorphism
group of the ${\cal N}=2$ $D=10$ type IIB SuperPoincar\'e algebra. Its $SO(2)$
subgroup is identified with a non-local $SO(2)$ transformation found in
hep-th/9806161. Performing T-duality, new non-local transformations are found 
in type IIA relating, among others, BIon configurations with two D2-branes 
intersecting at a point. Its M-theory origin is explained. These results show 
that part of the SuperPoincar\'e algebra automorphism group might be realized 
on the field theory as non-local transformations.
\\[1ex]
PACS numbers: 11.10.Kk, 11.25.-w\\
Keywords: Automorphisms, non-local symmetries, BPS states
\end{minipage}
}

\maketitle


\section{Introduction}

The relation among spacetime supersymmetry algebras and superbrane effective
actions is nowadays quite well understood \cite{sim-1}. The latter provide a 
field theory realization of the former. It then follows that there must exist,
to certain extent, a parallelism among a pure algebraic approach to M-theory 
(string theory) and a field theory approach on branes. In particular, BPS 
states can be algebraically characterized by the saturation of a BPS bound 
associated with the positivity of the matrix $<\alpha\vert \{Q_\alpha\, , 
Q_\beta\}\vert\alpha>\geq 0$ for all Clifford valued states $\vert\alpha>$
\cite{sim-2}. Such states do have a field theory description in terms of a 
special class of field configurations, the so called {\em BPS configurations}.
These saturate a bound on the field theoretic energy functional whenever 
certain functional constraints or {\em BPS equations}, are satisfied 
\cite{sim-3}. The latter can be derived \cite{sim-4bis} either 
from a direct hamiltonian analysis or from the resolution of the kappa 
symmetry preserving condition $(\Gamma_\kappa\epsilon=\epsilon)$ \cite{sim-4}
that any supersymmetric (bosonic) configuration must satisfy.

It has been lately stressed \cite{sim-5,sim-6,sim-6bis,sim-7} that the maximal 
automorphism group of the ${\cal N}=1$ $D=11$ SuperPoincar\'e algebra is 
$GL(32,\bR)$. This raises a natural question : is there any field theory 
realization for such automorphism group or a subgroup of it ?. In 
\cite{sim-5}, it was already pointed out the existence of an M5-brane symmetry
corresponding to one of such automorphism transformations. One of the purposes
of this paper is to go in this direction.

We shall concentrate on branes propagating in SuperPoincar\'e superspace.
Since the Lorentz group is a subgroup of the corresponding automorphism
group, it is already clear that such a subgroup will be linearly realized on 
the brane (before any gauge fixing). On the other hand, since ``central
charges'' appearing in maximal SuperPoincar\'e algebras admit a field theory
realization in terms of topological charges given by world space integrals
involving derivatives of the brane dynamical fields, and they are generically
``rotated'' under automorphism transformations \cite{sim-7}, one should also 
expect, if any, the existence of non-local transformations in the field theory
side.

We shall provide evidence for the existence of such non-local transformations.
We shall start by analysing the automorphism group of the  
${\cal N}=2$ $D=10$ type IIB SuperPoincar\'e algebra. It will be shown that
such group contains an $SL(2,\bR)$ factor, the corresponding U-duality group
in ten dimensional type IIB theory \footnote{It would be interesting to
know whether this relation among the automorphism group and the U-duality
group extends to lower dimensional superalgebras.}. We shall identify its
$SO(2)$ subgroup with the non-local $SO(2)$ transformations found in
\cite{sim-8} by analyzing them on some particular class of on-shell BPS 
configurations, dyons \cite{sim-3}. Performing a longitudinal T-duality 
transformation, we shall find new non-local transformations leaving the 
D2-brane effective action invariant. The new feature of the latter 
transformations is that they also involve bosonic scalar matter fields. The 
existence of these transformations again matches the corresponding $SO(32)$ 
automorphism transformations of the ${\cal N}=2$ $D=10$ type IIA 
SuperPoincar\'e algebra derived from a pure algebraic formulation
describing T-duality among type IIA and type IIB superalgebras \cite{sim-10}.
When applied on-shell, they map (among others) fundamental strings ending on 
the D2-brane (BIons) with two intersecting D2-branes at a point 
$(D2\perp D2(0))$. We end up with the M-theory interpretation of these new 
type IIA transformations. These are rotations (elements of $SO(10)\in SO(32)$)
involving the extra (eleventh) dimension. This interpretation is consistent 
with the well-known fact that an S-duality transformation in type IIB is a 
rotation interchanging the two independent cycles on the 2-torus in M-theory
\cite{sim-11}. From the field theory perspective (M2-brane effective action), 
the origin of the type IIA non-local transformations is the three dimensional 
world volume dualization \cite{sim-12}, needed to map the membrane action to 
the D2-brane action, which transforms the linearly realized rotations into 
the forementioned non-local ones. In this way, consistency of these new 
transformations with the known web of dualities in M/string theories is proved.

\section{S-duality, automorphisms and D3-branes}

The basic anticommutation relation defining the type IIB ${\cal N}=2$ $D=10$
SuperPoincar\'e algebra \footnote{We are using the same notation
and coventions as those used in \cite{sim-10}.} is given by
\bea
\{Q^i,Q^j\} &=& \cP^+\G^M Y_M^{ij} + \cP^+\frac{1}{3!}\G^{MNP}\epsilon^{ij}
Y_{MNP} \nonumber \\
& & + \cP^+\frac{1}{5!}\G^{M_1\dots M_5}Y_{M_1\dots M_5}^{+ij}\,,
\label{sim-algebra}
\eea
where
\bea
Y^{ij}_M &=& \delta^{ij}Y_M^{(0)} + \tau_1^{ij}Y_M^{(1)} +
\tau_3^{ij}Y_M^{(3)} \\
Y^{+ij}_{M_1\dots M_5} &=& \delta^{ij}Y_{M_1\dots M_5}^{+(0)} + 
\tau_1^{ij}Y_{M_1\dots M_5}^{+(1)} + \tau_3^{ij}Y_{M_1\dots M_5}^{+(3)} \,.
\eea

It would be important in the following to remember that all previous
charges appearing in the right hand side of equation (\ref{sim-algebra})
are associated with single $\nu=\frac{1}{2}$ BPS branes in certain
directions of spacetime. In particular, the three form spacelike components
$Y_{mnp}$ describe D3-branes standing along the mnp-hyperplane. Analogously,
the one form spacelike components $Y^{(1)}_m$ and $Y^{(3)}_m$ correspond
to D-strings and  fundamental strings stretching along the $x^m$ direction, 
respectively, whereas the five form spacelike components
$Y_{m_1\dots m_5}^{+(1)}$, $Y_{m_1\dots m_5}^{+(3)}$ and
$Y_{m_1\dots m_5}^{+(0)}$ describe D5-branes, NS5-branes and KK5B monopoles 
along the $m_1\dots m_5$-hyperplane, respectively. 

Just as for the ${\cal N}=1$ $D=11$ SuperPoincar\'e algebra, one could ask
about its maximal automorphism group. The latter certainly contains an
$SL(2,\bR)$ factor acting on the internal indeces $i,j$. If we consider
the transformation $\7Q^i = \left(U\,Q\right)^i$, $U\in \mbox{SL}(2,\bR)$, 
the latter will indeed be an automorphism of the algebra (\ref{sim-algebra})
if the charges on the right hand side transform as
\be
\7\cZ^{ij} = \left(U\cZ U^t\right)^{ij}\,.
\label{sim-trafo0}
\ee
Let us briefly study the elementary transformations generated by
$U_a=e^{\alpha\tau_a/2}$ $a=1,3$ and $U_2=e^{\alpha i\tau_2/2}$, where
$\{\tau_A \quad A=1,2,3\}$ is the set of Pauli matrices. Direct 
application of the transformation law (\ref{sim-trafo0}) determines the 
corresponding charge transformations. Since $U_a=U_a^t$, it can be checked
that $Y_M^{(0)}$ and $Y_M^{(a)}$ transform under an $SO(1,1)$ transformation
\begin{equation}
\left(\begin{array}{cl}
\7Y_M^{(0)}\\
\7Y_M^{(a)}
\end{array}\right) = S
\left(\begin{array}{cl}
Y_M^{(0)}\\
Y_M^{(a)}
\end{array}\right) 
\end{equation}
where
\begin{equation}
S= \left(\begin{array}{ccl}
\cosh\alpha & \sinh\alpha \\
\sinh\alpha & \cosh\alpha
\end{array}\right)\in \mbox{SO}(1,1)\,.
\end{equation}
There exists an analogous tranformation for the doublet
$\left(Y^{+(0)}_{M_1\dots M_5},\, Y^{+(a)}_{M_1\dots M_5}\right)$, all
other charges remaining invariant, due to the Pauli matrices algebra. 
Notice that $Y_0^{(0)}$ transforms, so that the energy
is not left invariant under such transformations. 

On the other hand, since $U_2^{-1}=U_2^t$, the subgroup generated by
$U_2(\alpha)$ transformations corresponds to the $SO(2)$ subgroup 
preserving the energy. In this case $Y_M^{(1)}$ and $Y_M^{(3)}$ transform as 
a doublet under the $SO(2)$ transformation
\begin{equation}
\left(\begin{array}{cl}
\7Y_M^{(1)}\\
\7Y_M^{(3)}
\end{array}\right) = R
\left(\begin{array}{cl}
Y_M^{(1)}\\
Y_M^{(3)}
\end{array}\right) 
\end{equation}
where
\begin{equation}
R= \left(\begin{array}{ccl}
\cos\alpha & \sin\alpha \\
-\sin\alpha & \cos\alpha
\end{array}\right)\in \mbox{SO}(2)\,.
\end{equation}
Analogous transformation properties are shared by the five form pair
$\{Y^{+(1)}_{M_1\dots M_5},\,Y^{+(3)}_{M_1\dots M_5}\}$. All
other charges remain invariant, including $Y_{mnp}$ and $Y^{+(0)}_{m_1\dots
m_5}$.

Notice that $U_2(\alpha)$ includes S-duality transformations. Indeed,
$U_2(\pi/2)$ interchanges D-strings with fundamental strings, and D5-branes 
with NS5-branes, whereas D3-branes and KK5B monopoles are left invariant,
in agreement with their known S-selfdual properties.

We shall concentrate in the following on $U_2(\alpha)$ transformations and 
their world volume realization. The natural brane effective action where one 
might look for such transformations is the D3-brane effective action. There
are two basic reasons for such choice. The first one is the breaking of the
automorphism group by the presence of the brane \cite{sim-6}. Due to the
S-duality covariance of the D3-brane effective action, this seems to be a
good choice. Furthermore, such classical action is known to admit solitonic
solutions corresponding to $(p,q)$-strings (dyons). It is then natural to
look for symmetry transformations relating $(p,q)$-strings with
$(p^\prime,q^\prime)$-strings.

We shall thus concentrate on D3-branes propagating in SuperPoincar\'e.
Its ${\cal N}=2$ supersymmetry and kappa invariant action is given by
\bea
S_{D3} &=&\int~d^4\sigma~\left(\cL_{DBI}~+~\cL_{WZ}\right)
\nonumber \\
\cL_{DBI}&=&-~\sqrt{-\mbox{det}\,(\cG_{\mu\nu}+\cF_{\mu\nu})}\,,
\nonumber \\
\cL_{WZ}&=&C_{(2)}~\cF~+~C_{(4)}\,,
\label{sim-d3action}
\eea
where
\bea
\cG_{\mu\nu}&=& \Pi^m_\mu\Pi^n_\nu\eta_{mn} \nonumber \\
\cF &=& d\,V + \Omega_3 \,,
\eea
are defined in terms of the supersymmetric invariant one form
\be
\Pi^m=dx^m + \bar\theta\G^m d\theta\,,
\ee
whose components are $\Pi^m_\nu=\partial_\mu x^m - \bar\theta\G^m\6_\mu\theta$
\footnote{We are using the same conventions for the Dirac matrices and forms 
as those given in \cite{sim-schwarz}.}, and
\be
\Omega_i= \bar\theta\,\,\hat\slPi\tau_i d\theta
\ee
where $\,\hat\slPi=\hat\Pi^m\G_m$ and $\hat\Pi^m=\Pi^m -\frac{1}{2}\bar\theta
\G^md\theta$. The RR-fields can be found in \cite{sim-cederwall}. It will
be important for us just to know that $C_{(2)}=\Omega_1$.

The transformations we are interested in were actually found in
\cite{sim-8}. In that paper, the action (\ref{sim-d3action}) was proven
to be invariant under the following set of transformations :
\bea
\delta x^m &=& 0 \label{sim-trafo1} \\
\delta\theta &=& \frac{\lambda}{2}\,i\tau_2\theta \label{sim-trafo2} \\
\delta F_{\mu\nu} &=& \lambda K_{\mu\nu}  \label{sim-trafo3} \\
\delta K_{\mu\nu} &=& -\lambda F_{\mu\nu} \label{sim-trafo4}
\eea
where $K_{\mu\nu}=-\frac{1}{2}\varepsilon_{\mu\nu\rho\sigma}\7K^{\rho\sigma}$
\footnote{$\varepsilon_{\mu\nu\rho\sigma}$ denotes the covariantly constant
antisymmetric tensor with indices raised and lowered by $\cG_{\mu\nu}$.} and
\be
\7K^{\rho\sigma}=\frac{1}{\sqrt{-\mbox{det}\,\cG}}\frac{\partial{\cal L}_{D3}}
{\partial F_{\rho\sigma}}\,.
\ee
One can check that
\bea
K_{\mu\nu}&=& -\frac{\sqrt{-\mbox{det}\,\cG}}{\sqrt{-\mbox{det}\left(
\cG+\cF\right)}}\left(\7\cF_{\mu\nu} + \cT\cF_{\mu\nu}\right) \nonumber \\
&& + C_{(2)\mu\nu}\,,
\label{sim-kmn}
\eea
where $\7\cF$ stands for the Hodge dual of the two form $\cF$ and
$\cT=\frac{1}{4}\cF_{\mu\nu}\7\cF^{\mu\nu}$.

Some comments are in order at this point. First of all, since it is the
gauge invariant field strength $F_{\mu\nu}$ that transforms linearly 
(\ref{sim-trafo3}), it is indeed true that the gauge potential $V$ transforms 
non-locally as stressed in the introduction. In fact such a transformation was
written down explicitly in the bosonic case in \cite{sim-itoh}. The only 
matter fields that transform are the fermionic scalars $\theta^i$ whose 
infinitesimal transformation coincides with the infinitesimal transformations 
of the corresponding supersymmetry generators $Q^i$ used in the algebraic
analysis.

To get a more physical intuition of the previous transformations and to further
relate them with the algebraic discussion, we shall evaluate them on a  
well-known BPS configuration : {\em dyons}. These are $\nu=\frac{1}{4}$
solitons representing $(p,q)$ strings ending on the D3-brane. They
correspond to the array 
\bea
\ba{cccccccccccl}
D3: &1&2&3&\_&\_&\_&\_&\_&\_& \nn \\
F1: &\_&\_&\_&4&\_&\_&\_&\_&\_& \nn \\
D1: &\_&\_&\_&4&\_&\_&\_&\_&\_& \,.
\ea
\eea
As all BPS configurations, they are characterized by some BPS equations
\bea
E^a &=& F_{0a} =\cos\alpha\,\6_a y 
\label{sim-d3soliton1} \\
B^a &=& \frac{1}{2}\epsilon^{abc}F_{bc} = \sin\alpha\, \delta^{ab}\6_b y 
\quad a,b,c=1,2,3 
\label{sim-d3soliton2}
\eea
where $y$ stands for the transverse excited scalar field (along the 4th
spacetime direction, as indicated in the previous array),
and some supersymmetry projection conditions
\bea
\Gamma_{0123}\, i\tau_2\epsilon &=& \epsilon 
\label{sim-susy1} \\
\Gamma_{0y}\left(\cos\alpha\,\tau_3 + \sin\alpha\,\tau_1\right)\epsilon &=&
\epsilon \,.\label{sim-susy2}
\eea
Equation (\ref{sim-susy1}) tells us our solution is describing a D3-brane
along directions 123, whereas (\ref{sim-susy2}) describes a non-threshold
bound state of strings ($\tau_3$ factor) and D-strings ($\tau_1$ factor).
Such an interpretation is further confirmed by computing its energy. The latter
is given by
\be
E_{BPS}=E_{D3} + \sqrt{\left(Y^{(3)}_4\right)^2 + \left(Y^{(1)}_4\right)^2}\,,
\label{sim-energyd3}
\ee
where $E_{D3}$ corresponds to the vacuum energy of an infinite planar D3-brane
along 123 directions, whereas the second factor corresponds to the energy
of a non-threshold bound state of strings and D-strings. The computation
of (\ref{sim-energyd3}) is entirely field theoretical, and in particular,
we can express $Y^{(a)}_4$ in terms of the worldspace integrals
\be
Y^{(3)}_4=\int_{D3}\, \vec{E}\cdot\vec\nabla y \quad , \quad
Y^{(1)}_4=\int_{D3}\, \vec{B}\cdot\vec\nabla y\,.
\label{sim-fieldcentral}
\ee
Equation (\ref{sim-energyd3}) matches the pure algebraic analysis result.
This would have been derived by solving the eigenvalue problem \cite{sim-7}
\[
[\Gamma^4\left(\tau_1\,Y^{(1)}_4+\tau_3\,Y^{(3)}_4\right) + \Gamma^{123}
i\tau_2\, Y_{123}]\vert\alpha>= E_{BPS}\vert\alpha>\,.
\]

Let us evaluate transformations (\ref{sim-trafo3})-(\ref{sim-trafo4}) for this
configuration. From the on-shell equalities $\sqrt{-\mbox{det}\,\cG}\7F^{0a}
=B^a$ and ${\sqrt{-\mbox{det}\,\cG}}\7F^{ab}=\epsilon^{abc}F_{0c}$,
one derives
\bea
\cT &=& \frac{F_{0a}B^a}{\sqrt{-\mbox{det}\,\cG}} \nonumber \\
\7F_{0a} &=& -\7F^{0b}\left(\delta_{ba} + \6_b y \6_a y\right)\,,
\eea
which allow us to find
\bea
K_{0a} &=& -B^a \nonumber \\
K_{ab} &=& \epsilon_{abc}F_{0c}\, ,
\eea
where we have used that 
\[
\sqrt{-\mbox{det}\left(\cG+\cF\right)}=1+\sin^2\alpha\delta^{ab}\6_a y 
\6_b y\,.
\] 
The latter lead to the well-known $SO(2)$ infinitesimal transformations
\bea
\delta E^a &=& -\lambda B^a \nonumber \\
\delta B^a &=& \lambda E^a\, ,
\label{sim-rot}
\eea
whose finite form is
\bea
E^{\prime a} &=& \cos\lambda\, E^a -\sin\lambda\, B^a = \cos (\alpha+\lambda)
\6_a y \nonumber \\
B^{\prime a} &=& \sin\lambda\,E^a +\cos\lambda\, B^a = \sin (\alpha+\lambda)
\6_a y \, .
\eea
These transformations show that indeed we are rotating the field theory
realization of the 'central charges' appearing in the supersymmetry algebra
(\ref{sim-fieldcentral}), so that we have indeed realized the forementioned 
automorphism as a non-local symmetry on the D3-brane effective action. 
Furthermore, as stressed in \cite{sim-7}, by fine tuning the parameter of the 
transformation $(\lambda)$, one can set one of the charges of the 
non-threshold bound state to zero, the above computation being a particular 
example of such behaviour \footnote{It is when one goes to the quantum theory,
that the $\mbox{SL}(2,\bR)$ automorphism group becomes an $\mbox{SL}(2,\bZ)$
group, in order to be consistent with charge quantization.}. We would like
to close this section by pointing out that the SO(2) rotation among the 
electric and magnetic fields could have been derived by requiring kappa
symmetry covariance. This is that the solution to $\Gamma_\kappa \epsilon
=\epsilon$ is mapped to the corresponding solution to 
$\Gamma^\prime_{\kappa^{\prime}} \epsilon^\prime=\epsilon^\prime$, where
$\Gamma^\prime_{\kappa^{\prime}}$ depends on the transformed fields and
$\epsilon^\prime=U_2 (\alpha)\epsilon$.

\section{T-duality and D2-branes picture}

In the previous section, we discussed some automorphisms of type IIB
SuperPoincar\'e algebra and in particular, the way its $SO(2)$ subgroup
was realized non-locally on the dynamical fields describing the D3-brane
effective action. It is natural to wonder about this symmetry structure
in type IIA, both algebraically and from the D2-brane field theory
perspective. We leave the M-theory interpretation for the next section.

It is already known the relation among type IIA and IIB SuperPoincar\'e
algebras. If a T-duality is performed along a spacelike direction $s$, one
may choose the supercharges to be related as follows
\be
Q^+=Q^2 \quad , \quad Q^-=\G_s Q^1\,,
\label{sim-tcharge}
\ee
where $Q^{\pm}$ are the type IIA supercharges. In this way, the previous 
$U_2(\alpha)$ automorphism can be rewritten as 
$U_s=e^{\alpha/2\,\G_s\G_{11}}$, which indeed belongs to $\mbox{SO}(32)$, 
the subgroup of type IIA automorphisms 
preserving energy. The latter statement can be straightforwardly derived
from the M-algebra analysis done in \cite{sim-7}. Notice that $\G_{11}$
is the ten dimensional chirality operator, so that $U_s(\alpha)$ can not
be interpreted as an spacetime rotation. From the type IIA superalgebra
perspective, it has to do with the freedom one has to make the choice
(\ref{sim-tcharge}) when relating both superalgebras under T-duality.
As discussed before, such automorphisms will ``rotate'' several doublets
of charges appearing in type IIA, while keeping some others invariant.
In particular, charges $\cZ_{sm}$ and $\cZ_m$ corresponding to D2-branes
and fundamental strings will form an $\mbox{SO}(2)$ doublet under
$U_s(\alpha)$ transformations. We refer the reader to \cite{sim-7} for such 
a discussion.

Once we apply a T-duality transformation, we loose spacetime covariance, but
it is clear that one could have performed a T-duality along a different 
spacelike direction $s^\prime$, so that it should be expected not just a 
single transformation but a set of them, $U_m=e^{\alpha_m/2 \G_m\G_{11}}$ 
$m=1,\dots ,9$ to be relevant in the T-dual description. This will be 
confirmed in the field theory analysis.

Let us move to the field theory perspective. In 
\cite{sim-joan,sim-kiyoshijoan}, the way longitudinal T-duality is realized 
on D-brane effective actions was studied, not only at the level of the action 
functional but also on its symmetry structure. The latter will be particularly
useful for us in order to derive the symmetry structure inherited by the 
D2-brane from its T-dual D3-brane. Since no bosonic scalar field $(x^m)$ 
transforms under $U_2(\alpha)$ (see equation (\ref{sim-trafo1})), there will 
be no compensating diffeomorphism transformation coming from the partial gauge 
fixing involved in the world volume realization of T-duality $(x^s=\rho)$
\footnote{Spacetime coordinates $\{x^m\}$ are splitted into $\{x^{\hat{m}},
x^s\}$ whereas world volume ones according to
$\{\sigma^\mu\}=\{\sigma^{\hmu},\rho\}$.}. 
Thus, we can directly study the double dimensional reduction of the 
transformation laws (\ref{sim-trafo2})-(\ref{sim-trafo4}). Let us start from
the fermionic sector. As shown in \cite{sim-kiyoshijoan}, the relation among
type IIA and type IIB fermions is given by
\be
\theta_+=\theta_2 \quad , \quad \theta_-=\Gamma_s\theta_1\,.
\label{sim-tfermion}
\ee
Just as for the supersymmetry generators, it is possible to rewrite 
(\ref{sim-trafo2}) as
\be
\delta\theta=\frac{\lambda}{2}\Gamma_s\Gamma_{11}\theta\,,
\label{sim-fermion}
\ee
where, as usual, $\theta=\theta_+ + \theta_-$, the subindex indicating
its ten dimensional chirality.

Since under T-duality, one of the gauge field components $(V_\rho)$ becomes
a world volume scalar in the T-dual theory $(\7x^s)$, it must be expected
to get non-local transformations not just for the reduced gauge field 
components $(V_{\hat\mu})$ but also for $\7x^s$. Double dimensional
reduction of transformations (\ref{sim-trafo3}) and (\ref{sim-trafo4}) gives
\bea
& \delta K_{\hmu\hnu} = -\lambda F_{\hmu\hnu} \quad , \quad
\delta F_{\hmu\hnu} = \lambda K_{\hmu\hnu} & \\
& \delta K_{\hmu\rho} = -\lambda \partial_{\hmu}\7x^s\quad , \quad
\delta \partial_{\hmu}\7x^s = \lambda K_{\hmu\rho} &
\eea 
whereas the remaining scalars do still remain invariant 
$(\delta x^{\hat{m}} = 0)$. In the latter expressions, 
$K_{\hmu\hnu}$ and $K_{\hmu\rho}$ are given by
\bea
K_{\hmu\hnu} &=& -\frac{1}{\sqrt{-\mbox{det}\,(\cG+\cF)}}\left[
\cG_{\hmu\hat\alpha}\cG_{\hnu\hat\beta}\epsilon^{\hat\alpha\hat\beta
\hat\delta}\Pi^s_{\hat\delta}  \right.\nonumber \\
& & \left. +\left(\frac{1}{2}\epsilon^{\hat\alpha\hat\beta\hat\delta}
\cF_{\hat\alpha\hat\beta}\Pi^s_{\hat\delta}\right)\cF_{\hmu\hnu}\right]
\nonumber \\
& & + \bar\theta\,\,\hat\slPi_{\hmu}\G_s\G_{11}\6_{\hnu}\theta - 
(\hmu \leftrightarrow \hnu)\,,  \label{sim-tk1} \\
K_{\hmu\rho} &=& -\frac{1}{\sqrt{-\mbox{det}\,(\cG+\cF)}}\left[
\frac{1}{2}\cG_{\hmu\hat\alpha}\epsilon^{\hat\alpha\hat\beta\hat\delta}
\cF_{\hat\beta\hat\delta}\right]  \nonumber \\
& & +\bar\theta\Gamma_{11}\partial_{\hmu}\theta\,.
\label{sim-tk2} 
\eea
It must be understood that all fields appearing in (\ref{sim-tk1})
and (\ref{sim-tk2}) are type IIA fields, furthermore, the 
$(\hmu \leftrightarrow \hnu)$ prescription just applies to the third line
in (\ref{sim-tk1}).

Notice that transformation (\ref{sim-tk1}) is manifestly non-covariant,
since it depends on the direction along which we perfom T-duality, whereas
(\ref{sim-tk2}) is totally covariant. In order to check whether the T-dual
D2-brane action has more non-local symmetries than the ones described
before, we shall compute the commutator of a ten dimensional Lorentz 
transformation $(\omega^{mn})$ and one of our new symmetry transformations 
$(\lambda)$ :
\be
[\delta\,,\tilde\delta]\theta=\lambda\omega^{sp}\Gamma_p\Gamma_{11}
\theta=\tilde\lambda^p\Gamma_p\Gamma_{11}\theta\, .
\ee
Due to the antisymmetry of the Lorentz parameter $\omega^{sp}$, $p$ is
definitely different from $s$. This shows that our three dimensional field 
theory has a larger set of non-local transformations \footnote{Similar
arguments can be applied to the remaining bosonic scalar fields on the
brane.} which can be obtained just by making covariant the previous ones,
\bea
&\delta\theta = \frac{\lambda}{2}\Gamma_m\Gamma_{11}\theta \,,
\label{sim-final0} & \\
& \delta K^m_{\hmu\hnu} = -\lambda^m F_{\hmu\hnu} \quad , \quad
\delta F_{\hmu\hnu} = \lambda^m K^m_{\hmu\hnu} & 
\label{sim-final1} \\
& \delta K_{\hmu\rho} = -\lambda^m \partial_{\hmu}x^m\quad , \quad
\delta \partial_{\hmu}x^m = \lambda^m K_{\hmu\rho} &
\label{sim-final2}
\eea 
where
\bea
K^m_{\hmu\hnu} &=& -\frac{1}{\sqrt{-\mbox{det}\,(\cG+\cF)}}\left[
\cG_{\hmu\hat\alpha}\cG_{\hnu\hat\beta}\epsilon^{\hat\alpha\hat\beta
\hat\delta}\Pi^m_{\hat\delta}  \right.\nonumber \\
& & \left. +\left(\frac{1}{2}\epsilon^{\hat\alpha\hat\beta\hat\delta}
\cF_{\hat\alpha\hat\beta}\Pi^m_{\hat\delta}\right)\cF_{\hmu\hnu}\right]
\nonumber \\
& & + \bar\theta\,\,\hat\slPi_{\hmu}\G^m\G_{11}\6_{\hnu}\theta - 
(\hmu \leftrightarrow \hnu)  \label{sim-tk1bis}
\eea
We would like to stress that such 'enhancement' of symmetry is typical of
T-duality and it is certainly not constrained to the particular construction 
used here.

Just as for the D3-brane case, we shall analyze the behaviour of some 
particular BPS configuration under these new transformations. We shall 
T-dualize the previous dyonic configuration along the direction 3. The T-dual
array is given by
\bea
\ba{cccccccccccl}
D2: &1&2&\_&\_&\_&\_&\_&\_&\_& \nn \\
F1: &\_&\_&\_&4&\_&\_&\_&\_&\_& \nn \\
D2: &\_&\_&3&4&\_&\_&\_&\_&\_& \,.
\ea
\eea
By the analysis done in \cite{sim-kiyoshijoan}, it is known that its BPS 
equations are described by the double dimensional reduction of equations 
(\ref{sim-d3soliton1}) and (\ref{sim-d3soliton2})
\bea
F_{0\hat{a}} &=& \cos\alpha\,\partial_{\hat{a}}y 
\label{sim-d2soliton1} \\
\epsilon^{\hat{a}\hat{b}}\partial_{\hat{b}}\7x^3 &=& \sin\alpha\, 
\delta^{\hat{a}\hat{b}}\partial_{\hat{b}}y \quad \hat{a},\hat{b}=1,2 
\label{sim-d2soliton2} \\
F_{12} &=& 0 \,.
\label{sim-d2soliton3}
\eea
The third equation states that there are no D0-branes being described
by our configuration as can be further confirmed by looking at the 
supersymmetry projection conditions
\bea
\Gamma_{012}\epsilon &=& \epsilon \\
\left(\cos\alpha\,\Gamma_{0y}\Gamma_{11} + \sin\alpha \, \Gamma_{03y}\right)
\epsilon &=& \epsilon
\eea
which are obtained from (\ref{sim-susy1})-(\ref{sim-susy2}) by direct 
application of the fermionic rules (\ref{sim-tfermion}). Notice that when 
$\alpha=0$, we recover the usual BIon describing a fundamental string ending 
on the D2-brane, whereas for $\alpha=\frac{\pi}{2}$, we recover the 
Cauchy-Riemann equations describing the intersection of two D2-branes at a 
point, $D2\perp D2(0)$. Both configurations are related to each other by 
application of transformations (\ref{sim-final1}) and (\ref{sim-final2}). 
Computing them when (\ref{sim-d2soliton1})-(\ref{sim-d2soliton3}) are 
satisfied or just by double dimensionally reducing equations 
(\ref{sim-rot}) we get
\bea
\delta \vec{E} &=& -\lambda\star\nabla\7x^3 \nonumber \\
\delta \left(\star\nabla\7x^3\right) &=& \lambda \vec{E} \,,
\label{sim-rot2}
\eea
where we are using the standard two dimensional calculus notation, that is,
$\vec\nabla=(\6_1\,, \6_2)$ and $\star\vec\nabla=(\6_2\,, -\6_1)$. Its
finite transformation is
\bea
\vec{E}^\prime &=& \cos\lambda\, \vec{E} -\sin\lambda\,
\star\vec\nabla\7x^3 \nonumber \\
&=& \cos\,(\alpha+\lambda)\,\vec\nabla y 
\nonumber \\
\star\vec\nabla\7x^{\prime 3} &=&
\sin\lambda\, \vec{E} +\cos\lambda\, \star\vec\nabla\7x^3 \nonumber \\
&=& \sin\,(\alpha+\lambda)\,\vec\nabla y
\eea
Thus, as expected, by fine tuning the global parameter $\lambda$, we 
interpolate between BIon configurations and $D2\perp D2(0)$ intersections.

The SO(2) rotation described by (\ref{sim-rot2}) fits with the supersymmetry
algebra picture. In this case, the charge carried by the fundamental string
is given by the worldspace integral
\be
\cZ_y=\int_{D2}\,\vec E \cdot \vec\nabla y \,,
\ee
whereas the charge carried by the second D2-brane admits the field theory
realization
\be
\cZ_{3y}=\int_{D2}\,\star\vec\nabla x^3 \cdot \vec\nabla y\,.
\ee
Thus we see that $\cZ_y$, $\cZ_{3y}$ are indeed rotated under (\ref{sim-rot2})
transformations, as the pure algebraic digression was suggesting to us.

\section{M-theory interpretation}

It is natural to wonder about the M-theory interpretation of any symmetry
structure found in type IIA theory. In the following, we shall reformulate all
previous results from an eleven dimensional perspective, both algebraically 
and from a field theory scenario. Due to the relation among ${\cal N}=1$ 
$D=11$ SuperPoincar\'e algebra and ${\cal N}=2$ $D=10$ type
IIA SuperPoincar\'e algebra, the latter being the dimensional reduction
of the former, it is straightforward to reinterpret the previous $SO(32)$ 
automorphisms as $SO(10)$ rotations, which we shall denote as
$U_m(\alpha) = e^{\alpha_m/2 \G_m \G_{\sharp}}$ 
\footnote{We use the symbol $\sharp$ 
to refer to the eleven dimensional direction, as it is usually done in the 
literature.}. Due to its rotational character, preservation of energy is
guaranteed. This picture also agrees with the well-known fact that an 
S-duality transformation $(\alpha=\frac{\pi}{2})$ in type IIB
is seen as a rotation interchanging the two independent cycles in the
two torus needed to relate M-theory with type IIB string theory 
\cite{sim-schwarz}.

The advantage of the M-theory formulation is that the previous non-local
transformations will be linearly realized on the M2-brane effective action.
It is actually very simple to match both results. Since the D2-brane effective
action is related to the M2-brane one by a world volume dualization
\cite{sim-12}, the linearly realized rotation $(\omega^{m\sharp})$ will 
induce a linear transformation on the gauge invariant quantity $\cF$, but a 
non-local one on the abelian $U(1)$ gauge field, as discussed previously.

Let us look into this connection more closely. Consider the three dimensional 
M2-brane effective action propagating in SuperPoincar\'e \cite{sim-esp}. 
The latter is invariant under the global $SO(10)$ rotations
\bea
\delta x^m &=& \omega^{my}\,y \nonumber \\
\delta y &=& -\omega^{my}\,x_m \nonumber \\
\delta\theta &=& \frac{1}{2}\omega^{my}\,\G_m\G_\sharp\, \theta\,,
\label{sim-m2trafo}
\eea
where $x_m=\eta_{mn}x^n$ and we have already splitted the eleven dimensional
bosonic scalar fields into $\{x^m,y\}$. The basic equation relating the scalar
field $y$ with its three dimensional dual $V$ is given by
\be
\6_\mu y -\5\theta\G_\sharp\6_\mu\theta = \frac{1}{2}\frac{v}{det\,\cG^{(10)}}
\cG^{(10)}_{\mu\nu}\epsilon^{\nu\alpha\beta}\cF_{\alpha\beta}\,,
\label{sim-rel}
\ee
where $v$ is an auxiliary scalar density whose value can be computed by 
solving its classical equation of motion
\be
v=\sqrt{-\mbox{det}\,\cG^{(11)}}=
\frac{-\mbox{det}\,\cG^{(10)}}{\sqrt{-\mbox{det}\,(\cG^{(10)}+\cF)}}\,.
\ee
Comparing with the objects appearing in the D2-brane transformations, we
realize that equation (\ref{sim-rel}) is equivalent to
\be
\6_\mu y = K_{\mu\rho}\, .
\label{sim-rel1}
\ee
{}From equation (\ref{sim-rel1}), we recover the set of transformations for the
dynamical fields on the D2-brane,
\bea
& \delta\6_\mu y=-\omega^{my}\6_\mu x_m \Leftrightarrow
\delta K_{\mu\rho} =-\omega^{my}\6_\mu x_m & \\
& \delta \6_\mu x^m = \omega^{my}\6_\mu y \Leftrightarrow 
\delta \6_\mu x^m = \omega^{my}K_{\mu\rho} & \,,
\eea
whereas the fermionic transformations are trivially identified since the
eleven dimensional Majorana spinors $\theta$ are splitted into
$\theta_1 + \theta_2$, the two different chiral Majorana-Weyl spinors
in type IIA. Thus the linear transformations (\ref{sim-m2trafo}) are
mapped, through the world volume dualization (\ref{sim-rel}), to
non-local transformations on the D2-brane action, by identifying
$\omega^{my}=\lambda^m$. Notice that this eleven dimensional
interpretation again agrees with the symmetry enhancement found in the
D2-brane when looking at it from its T-dual D3-brane.

To finish up the M-theory discussion, we shall comment on the uplifted
BPS configuration corresponding to the previous section. This is described
by the array
\bea
\ba{cccccccccccl}
M2: &1&2&\_&\_&\_&\_&\_&\_&\_&\_ & \nn \\
M2: &\_&\_&\_&4&5&\_&\_&\_&\_&\_ & \nn \\
M2: &\_&\_&3&4&\_&\_&\_&\_&\_&\_ & \,.
\ea
\eea
We shall set the static gauge $x^\mu=\sigma^\mu$ $\mu=0,1,2$ to describe
the location of the world volume brane and excite three transverse scalars
$x^i$ $i=3,4,5$. We shall find the BPS equations for this configuration by
solving the kappa symmetry preserving condition,
\be
\frac{1}{3!}\epsilon^{\mu\nu\rho}\6_\mu x^m \6_\nu x^n \6_\rho x^p
\Gamma_{mnp}\epsilon = \sqrt{-\mbox{det}\,\cG}\,\epsilon\,.
\label{sim-e1}
\ee

Just from inspection of the latter array, three single brane projectors will
be involved in the solution $\{\Gamma_{012},\Gamma_{045},\Gamma_{034}\}$.
Following the analysis presented in \cite{sim-7}, the supersymmetry
projection conditions must be
\bea
\Gamma_{012}\epsilon &=& \epsilon \label{sim-e2} \\
\left(\cos\alpha\,\Gamma_{045} + \sin\alpha\,\Gamma_{034}\right)
\epsilon &=& \epsilon \,. \label{sim-e3}
\eea
The left hand side of equation (\ref{sim-e1}) equals
\[
\left(\Gamma_{012} + \epsilon^{0ab}\6_b x^i \Gamma_{0ai} +
\frac{1}{2}\epsilon^{0ab}\6_a x^i \6_b x^j \Gamma_{0ij}\right)\epsilon
\]
where $a,b=1,2$. Requiring $\epsilon^{0ab}\6_b x^i \Gamma_{0ai}\epsilon$
to vanish and using the identities
\bea
\Gamma_{014}\epsilon &=& -\left(\cos\alpha\,\Gamma_{025} + \sin\alpha\,
\Gamma_{023}\right)\epsilon \nonumber \\
\Gamma_{024}\epsilon &=& \left(\cos\alpha\,\Gamma_{015} + \sin\alpha\,
\Gamma_{013}\right)\epsilon\,,
\eea
derived from (\ref{sim-e2}) and (\ref{sim-e3}), we get the set of BPS equations
\bea
\cos\alpha\,\vec \nabla x^4 &=& \star \vec\nabla x^5 
\label{sim-e4} \\
\sin\alpha\, \vec\nabla x^4 &=& \star \vec\nabla x^3 \,. \label{sim-e5} 
\eea
Notice that they interpolate among $M2\perp M2(0)$ configurations in definite
directions for $\alpha=0,\frac{\pi}{2}$. The remaining non-trivial
term $\frac{1}{2}\epsilon^{0ab}\6_a x^i \6_b x^j \Gamma_{0ij}$ splits
into
\be
\frac{1}{2}\epsilon^{0ab}\6_a x^{\hat{\i}} \6_b x^{\hat{\j}} 
\Gamma_{0\hat{\i}\hat{\j}} +
\frac{1}{2}\epsilon^{0ab}\6_a x^5 \6_b x^3 \Gamma_{053}\,.
\ee
The last one is identically zero due to (\ref{sim-e4}) and 
(\ref{sim-e5}), whereas the first one equals
\be
(x^4)^2\left(\cos\alpha\,\Gamma_{045} + \sin\alpha\,\Gamma_{034}\right)
\epsilon=(x^4)^2 \epsilon
\ee
by (\ref{sim-e4})-(\ref{sim-e5}) and (\ref{sim-e3}). Thus the left hand side 
equals $(1+(x^4)^2)\epsilon$.

To check the matching with the right hand side, we must compute the world
volume induced metric when (\ref{sim-e4}) and (\ref{sim-e5}) are satisfied. 
This is given by
\bea
& \cG_{00}=-1 \quad , \quad \cG_{0a}=\cG_{12}=0 & \\
& \cG_{11}=1 + \sum_i (\6_1 x^i)^2= 1 + (x^4)^2 & \\
& \cG_{22}=1 + \sum_i (\6_2 x^i)^2= 1 + (x^4)^2 &
\eea
and indeed shows that $\sqrt{-\mbox{det}\,\cG}\,\epsilon=(1+(x^4)^2)\epsilon$,
matching our previous computation.

At this point, one can check the existence of an $SO(10)$ rotation relating
the latter BPS configuration with
\bea
\ba{cccccccccccl}
M2: &1&2&\_&\_&\_&\_&\_&\_&\_&\_ & \nn \\
M2: &\_&\_&\_&4&5&\_&\_&\_&\_&\_ & \,.
\ea
\eea
Following our general discussion, such a rotation must be 
$U=e^{\beta\Gamma_{53}/2}$. Indeed, this rotation in the 35-plane transforms 
the BPS equations (\ref{sim-e4})-(\ref{sim-e5}) into
\bea
\star \vec\nabla x^{\prime 5} &=& \cos (\alpha+\beta)\, \vec\nabla x^4 \\
\star \vec\nabla x^{\prime 3} &=& \sin (\alpha+\beta)\, \vec\nabla x^4
\eea
which show that by setting $\beta=-\alpha$, $x^{\prime 3}$ becomes
constant, and there is thus no longer an excited scalar in that direction. 
The second supersymmetry condition (\ref{sim-e3}) is also conveniently 
mapped into $\Gamma_{045}\epsilon^\prime =\epsilon^\prime$, confirming our 
previous interpretation.

Rotations in the 35-plane indeed rotate the corresponding M2-brane
charges in the supersymmetry algebra, since they are given by
\bea
\cZ_{45} &=& \int_{M2}\, \star\vec\nabla x^5\cdot \vec\nabla x^4 \,,
\nonumber \\
\cZ_{34} &=& \int_{M2}\, \star\vec\nabla x^3\cdot \vec\nabla x^4\,.
\eea

Once we have understood the M-theory configuration, it is straightforward
to recover all previous D2-brane BPS equations (\ref{sim-d3soliton1})-
(\ref{sim-d3soliton2}) and supersymmetry conditions (\ref{sim-susy1})-
(\ref{sim-susy2}) by explicitly using the relation (\ref{sim-rel1})
on-shell. This finishes the consistency check of the exposed non-local 
transformations into the web of M/string theory dualities.


{\em Acknowledgements}

JS was supported by a fellowship from Comissionat per a Universitats i 
Recerca de la Generalitat de Catalunya and is presently being supported
by a fellowship from the Feinberg Graduate School. This work was supported 
in part by AEN98-0431 (CICYT), GC 1998SGR (CIRIT).


\end{document}